\documentclass[aps,prx,twocolumn,superscriptaddress,nofootinbib]{revtex4-2}

\usepackage{amsmath,amssymb,amsfonts}
\usepackage{graphicx}
\usepackage{booktabs}
\usepackage{xcolor}
\usepackage{siunitx}
\usepackage{listings}
\usepackage[colorlinks=true,allcolors=blue]{hyperref}

\newcommand{\todo}[1]{\textcolor{red}{\textbf{[TODO: #1]}}}
\newcommand{\note}[1]{\textcolor{teal}{\textsf{[#1]}}}
\renewcommand{\todo}[1]{}
\renewcommand{\note}[1]{}

\newcommand{\mc}{\ensuremath{\mathrm{max\_corrected}}}
\newcommand{\dist}{\ensuremath{d}}
\newcommand{\pL}{\ensuremath{p_{L}}}
\newcommand{\nq}{\ensuremath{n_{q}}}
\newcommand{\Gsim}{\ensuremath{G_{\mathrm{sim}}}}
\newcommand{\Gint}{\ensuremath{G_{\mathrm{int}}}}

\begin{document}

\title{Evolving Quantum Error-Correcting Encodings for Molecular Simulation}

\author{Kenny Heitritter}
\affiliation{qBraid Co.}
\author{James Brown}
\affiliation{qBraid Co.}
\author{Tarini Hardikar}
\affiliation{qBraid Co.}

\date{\today}

\begin{abstract}
Useful quantum algorithms require many coupled discrete design choices. We
study LLM-driven evolutionary program synthesis---a language model edits a
program, an external verifier scores the result, and high-scoring programs are
retained and re-mutated---as a tool for quantum-computing research. As a case
study, we apply this loop to the Generalized Superfast Encoding (GSE), a
fermion-to-qubit encoding whose prior molecular constructions reach code
distance $3$. The search discovered interpretable constructor programs whose
codes have \emph{exact} distance $5$ on the molecular instances tested, and
distance $6$ on one $20$-mode instance, under strict stabilizer-coset
semantics. To our knowledge these are the first GSE/superfast encodings beyond
distance $3$ for dense molecular Hamiltonians. A second search, guided by
verifier analysis of the first artifact, found a circulant constructor that
reaches a five-qubits-per-mode floor on the tested $12$-, $14$-, $16$-, and
$20$-mode instances, with certified dense-rule fallback at the failing
$18$-mode case. As secondary resource descriptors, in a code-capacity
\emph{memory} comparison at $p=10^{-3}$ the resulting encodings use
$4.2$--$5.0\times$ fewer data qubits than a scoped per-mode Jordan--Wigner
$+$ $[[25,1,5]]$ surface route and have $3.4$--$8.2\times$ lower
logical-failure rates under finite-weight decoding tables with explicit
truncation brackets; we claim no circuit-level fault-tolerance or Trotter-cost
advantage. The search trajectory illustrates a general operating lesson:
rewarding distance alone selects trivial dense graphs, whereas holding verified
distance fixed and rewarding compression selects structured rules.
\end{abstract}

\maketitle

\section{Introduction}
\label{sec:intro}

% TODO(headline figure): commission a professionally-edited headline figure
% for page 1. Intended content: AlphaEvolve/ShinkaEvolve program score vs.
% evaluation index, with a few annotated waypoints captioning -- at a high
% level -- the program the system had written at that point (distance-1 seed
% -> dense complete-graph distance climb -> fixed-distance gate -> 2-factor
% deletion / compression -> 168-qubit winner). This is the polished redo of
% the rough trajectory plot in App.~\ref{app:climb} (Fig.~\ref{fig:trajectory}),
% which is NOT yet headline quality. Drop the \begin{figure*}...\end{figure*}
% here once the graphic is ready.

Making quantum computers useful is as much a \emph{design} problem as a
hardware problem. Between an abstract algorithm and a physical device sits a
stack of choices: how to encode the problem into qubits, which
error-correcting structure to carry, how to compile and decode. Optimizing each
is a search over a combinatorially large space of coupled, discrete
alternatives whose resource cost decides feasibility. Today these choices tend
to be made almost entirely by hand. Where automation exists at
all, as it does for the fermion-to-qubit encoding studied
here~\cite{simkovic2024,yu2025}, it optimizes \emph{within} a
predetermined parametric family.

Large language models (LLMs) enable evaluator-guided search in which the mutable object is a \emph{program} rather than a parameter vector. In this \emph{evolutionary program synthesis}, a language model proposes edits to a seed program; an immutable evaluator scores each mutated candidate; and high scorers are retained and re-mutated. FunSearch demonstrated the recipe on extremal combinatorics in 2023~\cite{romeraparedes2024}; successors have scaled it from
outperforming top human teams in combinatorial programming
contests~\cite{velickovic2024} to recovering stranded compute in
planet-scale cluster scheduling and improving matrix-multiplication
kernels~\cite{novikov2025}. Because the output is source code, it can be inspected as a proposed rule rather than only as a point in parameter space. The evaluator determines the search objective, so the returned program is machine-proposed but goal-directed by a fixed external measurement.

How researchers should operate such systems remains unsettled. A recent
study of mathematicians working with AlphaEvolve, an evolutionary coding
agent, finds that the work divides between deciding what to ask the system
for and making sense of what it returns, and argues that such systems are
best treated as scientific instruments rather than as
assistants~\cite{bauerle2026}. That framing matches our experience
throughout the present work (\S\ref{sec:disc-lessons}). We use this case study, based on early access to AlphaEvolve~\cite{novikov2025} and an open-source replication~\cite{lange2025}, to extract practical lessons for applying evaluator-guided program synthesis to quantum-computing design problems.

In this work, we use the search to push the error-correcting power of the Generalized Superfast Encoding (GSE)~\cite{bravyi2002,setia2018,brown2025}, a fermion-to-qubit encoding with inherent error-correction properties.  The fermion-to-qubit encoding sets the entry cost of quantum chemistry on quantum hardware: it fixes the qubit count, the Hamiltonian term weight (and hence Trotter-circuit depth) and encoding distance. The encoding distance is the fewest number of single-qubit operations that map one encoded state to another. If the code carries real distance, the logical error rate is another important property to examine. Prior GSE constructions for molecules have only reached distance 3 which we aim to expand.

The contribution in this case study is deliberately scoped. The evaluator-guided search finds constructor-level GSE rules whose outputs have exact distance 5 on the molecular instances tested, and distance 6 for one dense 20-mode instance, under strict stabilizer-coset semantics---to our knowledge the first GSE/superfast encodings beyond distance 3 for dense molecular Hamiltonians. As secondary descriptors, in a code-capacity \emph{memory} comparison these codes additionally use fewer data qubits and have lower logical-failure rates than a textbook per-mode JW$+[[25,1,5]]$ surface route. We do not claim a circuit-level fault-tolerance advantage or a lower Trotter-step error; both require syndrome-extraction and encoded-operation analyses beyond the present work. The main methodological point is that the verifier-grounded search produced readable constructors, not just isolated code instances, and that the useful constructors appeared only after the objective was staged as fixed-distance compression.

\section{The design problem}
\label{sec:background}

\subsection{Fermionic encodings with intrinsic protection}
\label{sec:bg-gse}
Simulating a fermionic Hamiltonian on qubits requires a fermion-to-qubit
encoding: a map from creation/annihilation operators to Pauli operators
preserving the canonical anticommutation relations. The textbook
Jordan--Wigner (JW) transform uses one qubit per mode but produces Pauli
strings whose weight grows linearly in the mode index, and high-weight
operators are the expensive ones: exponentiating a weight-$w$ term in a
Trotter step costs an entangling-gate ladder of depth $O(w)$.
Bravyi--Kitaev and ternary-tree
encodings~\cite{bravyi2002,jiang2020} reduce the weight to $O(\log n)$, but
all of these are \emph{distance-1}: a single qubit error maps to a single,
undetectable fermionic error. Protection can be added on by concatenating
a generic code on top, multiplying the qubit count by the outer block size.

\begin{figure}[t]
\centering
\includegraphics[width=0.72\columnwidth]{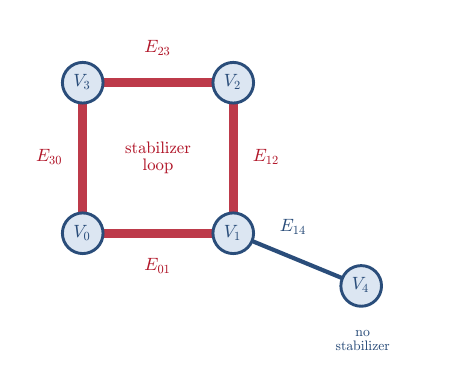}
\caption{Where a GSE encoding's protection comes from. Each vertex of
$\Gsim$ carries a vertex operator $V_i$ (an encoded mode-parity
observable, i.e.\ a \emph{logical} operator); each edge an edge operator
$E_{ij}$ (an encoded hopping). The product of edge operators around any
closed loop---here $S=E_{01}E_{12}E_{23}E_{30}$---is a \emph{stabilizer
generator}; the pendant edge $E_{14}$, on no cycle, contributes none. A
connected graph yields $|E|{-}|V|{+}1$ independent generators
(\S\ref{sec:dist-terminology}).}
\label{fig:gse-schematic}
\end{figure}

The superfast and Generalized-Superfast (GSE)
encodings~\cite{bravyi2002,setia2018,brown2025,steudtner2019,derby2021,jiang2019}
instead place the fermionic
degrees of freedom on the edges and vertices of a \emph{simulation graph}
$\Gsim$, representing fermionic operators as products of edge operators
$E_{ij}$ and vertex operators $V_i$, so Pauli weight is governed by graph
path-lengths rather than a global ordering. The graph's cycle structure
induces \emph{stabilizer generators}---products of edge operators around
closed loops that act as the identity on the code space
(Fig.~\ref{fig:gse-schematic})---giving the family \emph{intrinsic} error
detection and, with rich enough connectivity, correction. Published results
reach distance 3 analytically at interaction-graph vertex degree
${\ge}6$~\cite{setia2018} and an error-\emph{detecting} (distance-2) molecular
variant~\cite{brown2025}, with higher distance stated as a goal (App.~\ref{app:semantics}).

We define four terms before proceeding. The \emph{complement} of a graph $G$
has the same vertices as $G$ and an edge exactly where $G$ does not. A
\emph{Hamiltonian cycle} is a closed walk visiting every vertex exactly
once. A \emph{2-factor} is a 2-regular spanning subgraph---a disjoint set of
cycles that together cover every vertex; a Hamiltonian cycle is the
connected special case, and we name a 2-factor by its cycle-length type
(e.g.\ $6{+}4{+}4$). A \emph{circulant graph} $C_n(s_1,\dots,s_k)$ places
$n$ vertices on a ring and joins each vertex to those $s_1,\dots,s_k$
steps away in both directions.

Constructing a GSE encoding for a molecule means choosing three coupled
objects. The first is the \emph{simulation graph} $\Gsim$. It may add ancilla
vertices and edges beyond the bare interaction graph $\Gint$, and it dominates
every structural metric. A mode on a vertex of valence $v$ occupies
$\lceil v/2\rceil$ qubits, and the cycle space fixes the stabilizers and hence
the achievable distance. The second is the \emph{mode--vertex map}, an injection
that decides which encoded operators are short and which are long. The third is
the \emph{operator type}, the Pauli convention per Majorana operator. It can be
any Clifford-rotated variant, and it is by far the largest axis. For an eight-mode
molecule the joint space exceeds $10^{50}$, far beyond
enumeration,\footnote{For H4 padded to $11$ vertices:
$2^{\binom{11}{2}}\!\approx\!4\times10^{16}$ simulation-graph edge subsets,
$11!/3!\approx 7\times10^{6}$ mode-to-vertex injections, and the
operator-convention axis, where the Clifford group on $n$ qubits has order
$2^{n^{2}+2n}\prod_{j=1}^{n}(4^{j}{-}1)\approx 8\times10^{26}$ at $n{=}6$
qubits per mode. One shared convention already gives
${\gtrsim}2\times10^{50}$; conventions varying across the vertex sizes that
occur, multi-edges, and ancilla-count choices push the count well past
$10^{80}$.} so
what matters is not coverage but \emph{navigation}.

\subsection{Distance and verification semantics}
\label{sec:dist-terminology}
We quantify protection by $\mc$, the largest $t$ such that \emph{every} Pauli
error of weight $\le t$ produces a unique syndrome; $\mc{=}t$ certifies code
distance $\dist\ge2t{+}1$, conservatively for degenerate codes. All claims
use \emph{strict} stabilizer-coset semantics (the vertex operators $V_i$
are encoded logical observables, so an error equal to a product of $V_i$
counts as a failure), under the single equivalence relation stated
canonically in App.~\ref{app:semantics}, the stabilizer-group definition
under which the original distance-3 theorem was proved~\cite{setia2018}.
Where we state a distance as \emph{exact}, exhaustive enumeration found
zero logical operators below it and exhibited the minimum-weight logicals
at it. Logical failure rates $\pL(p)$ are measured in the code-capacity
model with the same semantics using finite-weight minimum-weight decoder
tables and explicit truncation brackets; we use them to compare memory routes,
not to predict hardware performance (\S\ref{sec:disc-limits}).

\section{Method}
\label{sec:methods}

The method is the evolutionary program synthesis of
\S\ref{sec:intro}~\cite{romeraparedes2024,novikov2025}: maintain a
population of candidate \emph{programs}; have a language model propose
source-level edits to a designated mutable region; score every candidate
through an immutable evaluator---here an exact stabilizer verifier---and
retain and re-mutate high scorers. The rest of this section specializes
that loop to the Generalized Superfast Encoding: the evolved object, the evaluation
panel, the staged objective, the seeds, the apparatus, and the defenses.

\paragraph{The evolved object.}
\label{sec:methods-object}
The search target is a Python function, not a fixed code:
\texttt{build\_encoding(\Gint, fer\_op) -> EncodingSpec(\Gsim,
mode\_vertex\_map, op\_type)}.
It receives the molecule's interaction graph and fermionic operator and
returns the three coupled axes of \S\ref{sec:bg-gse}. The evaluator wraps
the returned specification with the upstream GSE constructor, a trusted,
non-evolved backbone whose validity check (the fermionic anticommutation
relations) is a hard gate. All three axes live in a single mutable region;
the scaffolding (imports, graph and operator helpers, a Clifford-rotated
operator factory) is fixed, and the molecule's identity is withheld, so the
search must express structural rules rather than per-molecule lookups.

\paragraph{Panel, verification, and baselines.}
\label{sec:methods-eval}
The search optimizes over a training panel of three molecules at sto-3g (H4,
H6, LiH; 8/12/12 spin-orbitals); two held-out molecules (BeH2, H2O; 14
modes) are never exposed during the search and are evaluated only
afterwards, on the top-ranked programs. Post-search experiments apply the
\emph{frozen} constructors to further molecules (NH3, CH4, N2;
\S\ref{sec:results-family}); the search panel itself never grew. Every reported $\mc$ is recomputed by an exact
strict-coset verifier (deterministic, no statistical uncertainty); final
distances are pinned \emph{exact} by exhaustive minimum-weight-logical enumeration, and logical-failure rates use the same finite-weight minimum-weight decoder construction on both routes, with truncation brackets carried explicitly (App.~\ref{app:eval}). The
conventional QEC baseline concatenates \emph{textbook} Jordan--Wigner
(exactly $n_{\rm modes}$ logical qubits) with an independent
distance-matched block per logical qubit, the molecule failing if any block
fails. The headline variant uses one $[[25,1,5]]$ rotated surface code per
mode---we call this whole pipeline the \emph{surface route} throughout---and
we also report the $[[17,1,5]]$ color-code variant. A hand-designed sweep (B1, App.~\ref{app:eval}) and a
matched-compute random search over the same axes (B2,
\S\ref{sec:results-climb}) establish what hand design and unguided
sampling reach.

\paragraph{A climb-then-compress pipeline.}
\label{sec:methods-objective}
The methodological core is a two-stage pipeline on the \emph{same} encoder
family, panel, and apparatus. Both stages use additive,
resource-normalized scores sharing one \emph{resource term} (a
logarithmic penalty on excess qubits, term weight, and stabilizer weight
relative to a conventional reference; App.~\ref{app:objectives}) and
differ only in how distance enters.
\textbf{Stage 1 (climb; Regime~A)} treats \emph{distance as reward}: each
molecule contributes a smooth distance ladder---the strict fraction
$d_k\in[0,1]$ of weight-exactly-$k$ errors corrected, with geometrically
growing weights---so saturating a new distance tier dominates everything
else [Eq.~\eqref{eq:regimeA}]. Its job is discovery; its output is a
\emph{verified, over-provisioned high-distance seed}.
\textbf{Stage 2 (compress; Regime~B)} treats \emph{distance as a held constraint}: a hard gate requires the candidate to hold the target strict distance on every molecule; above it the score is a fixed bonus plus the resource term [Eq.~\eqref{eq:regimeB}], so accepted improvements come only from reducing resources while preserving distance. The
compression gate presupposes the verified seed that the climb supplies.
Why the climb alone is not enough is shown in \S\ref{sec:results-climb}:
its best reached basin is a resource-blind dense graph, which under the gate
becomes the worst gate-passing code. The remaining alternative, a
\emph{single-stage} objective that shapes distance reward and unlocks the
compression term at the target, launched directly from the distance-1
seed, was tested as a matched-budget control: it reaches distance 5 but
not the compression (\S\ref{sec:disc-lessons}).

\paragraph{Seeds.}
\label{sec:methods-seed}
The climb starts from a genuine \emph{distance-1} cell (a path graph: no
cycles, no stabilizers, $\mc{=}0$ everywhere). The compression stage starts
from a verified but deliberately \emph{over-provisioned} code---the leanest
strict distance-5 program from a prior climb, holding $\mc{=}2$ on all
three molecules at $272$ panel qubits with visible redundancy---and runs
at fixed targets distance-3 (control) and distance-5 (flagship). Each stage
is paired with its natural seed, so
the controlled comparison is between \emph{regimes} (objective plus matched
seed), not the reward variable in isolation.

\paragraph{Search apparatus and cross-apparatus replication.}
\label{sec:methods-apparatus}
Two search apparatuses produced the distance-5 compression result. The
discovery runs used Google's AlphaEvolve Cloud API (early-access
program)~\cite{novikov2025}, with mutations proposed by a mixture of Gemini
3.1 Pro and Gemini 3 Flash; its fixed-distance $d{=}5$ run evaluated $1411$
candidates and compressed the $272$-qubit seed panel to $168$. The
replication used the open-source ShinkaEvolve~\cite{lange2025} driven by
GPT-5.5: starting from the \emph{same} $272$-qubit seed, it re-derived the same 2-factor-deletion rule and
the same $168$-qubit panel under a different apparatus and LLM family. The floor-family run (\S\ref{sec:results-floor}) and the
stabilizer-basis searches (\S\ref{sec:disc-limits}) also used ShinkaEvolve,
driven by the cheaper GPT-5.4 under hard \$$20$ budgets. The caps did not
bind uniformly: the basis search's best program appeared ${\sim}\$6$ in,
with no improvement over the final $70\%$ of its budget, while
the joint search was still improving at its cap, yet its
surviving open case withstood a doubled-budget, stronger-model
continuation and was then closed by an exact optimality certificate
(\S\ref{sec:disc-limits}). Budgets, stop
conditions, and all search
configurations, seeds, evaluators, scoring code, program dumps, and raw outputs
are retained in qBraid's internal experiment records. They are not a public
reproduction package: the current evaluator and constructors depend on qBraid's
closed-source GSE implementation, and the source snippets printed here are meant
to document the discovered rules rather than to run stand-alone. Total paid LLM
spend across every search in this paper is under \$$250$
(App.~\ref{app:provenance}).

\paragraph{Defenses.}
\label{sec:methods-defenses}
Because the search optimizes exactly what the evaluator measures, we use
explicit evaluator defenses: AST scans, recomputation of every score by the exact
verifier, deterministic evaluation, scoring logic the model never sees, and
a never-exposed held-out panel (App.~\ref{app:evaluator}). These prevent
\emph{metric tampering}; they cannot prevent \emph{specification
gaming}~\cite{amodei2016,skalse2022,krakovna2020}---legitimately maximizing
a correctly-measured but under-specified objective---the failure mode the
climb stage exhibits (\S\ref{sec:results-climb}). A hard per-candidate
wall clock ($600$\,s in the discovery runs) backstops resource exhaustion
by runaway candidates; its price is a bias against slow-but-valid
constructors, a limitation we flag rather than solve.

\section{Results}
\label{sec:results}

\subsection{The discovered rule}
\label{sec:results-constructor}
The primary output is a compact constructor;
Fig.~\ref{fig:constructor-source} reproduces its mutable region verbatim.
Its central move: build a near-complete simulation graph on
$\max(n_{\rm modes},10)$ vertices, then \emph{remove a Hamiltonian cycle drawn from the complement of the interaction graph} $\Gint$. Removing a
2-regular subgraph lowers every vertex's valence by exactly two, so the GSE
qubit budget $\lceil\deg_{\max}(\Gsim)/2\rceil$ per vertex drops by one
(e.g.\ $6\to5$ on the 12-vertex graphs), while drawing the removed edges
from $\Gint$'s \emph{complement} keeps every real hopping edge present and
short, so term weight is not inflated. The rule needs only the plain
distance-3 operator basis; no Clifford rotation is required for distance 5. Top-ranked variants also weight graph edges by the fermionic coefficients or vary the mode assignment. The displayed line count is specific to the AlphaEvolve panel winner: the ShinkaEvolve replication reaches the same
$168$-qubit panel through a $293$-line mutable region that generalizes the
deletion to a minimum-cost 2-factor dynamic program; that generalization is
what transfers to larger molecules (\S\ref{sec:results-efficiency}).

\begin{figure}[t]
\begin{lstlisting}[language=Python,
    basicstyle=\fontsize{6.0}{6.9}\selectfont\ttfamily,
    breaklines=true, breakindent=1.5em, columns=fullflexible,
    commentstyle=\itshape, showstringspaces=false]
n_sim = max(n_modes, 10)
G_sim = nx.complete_graph(n_sim)

if n_sim == 12:
    # To minimize Hamiltonian term weight, we remove a 12-cycle that lies entirely
    # in the complement of G_int. This ensures interactions remain direct edges.
    comp_G = nx.Graph()
    comp_G.add_nodes_from(range(12))
    for u, v in nx.non_edges(G_int):
        comp_G.add_edge(u, v)

    def find_cycle(path):
        if len(path) == 12:
            if comp_G.has_edge(path[-1], path[0]):
                return path
            return None
        # Warnsdorff's heuristic for fast cycle finding
        neighbors = [n for n in comp_G.neighbors(path[-1]) if n not in path]
        neighbors.sort(key=lambda x: sum(1 for nn in comp_G.neighbors(x) if nn not in path))
        for neighbor in neighbors:
            res = find_cycle(path + [neighbor])
            if res:
                return res
        return None

    cycle = find_cycle([0])
    if cycle:
        G_sim.remove_edges_from([(cycle[i], cycle[(i+1)%12]) for i in range(12)])
    else:
        G_sim.remove_edges_from(nx.cycle_graph(12).edges())

elif n_sim == 10:
    # Drop the edge between the two ancillae to reduce their valence to 8 (qpm=4)
    G_sim.remove_edge(8, 9)

mode_vertex_map = {i: i for i in range(n_modes)}

# Ultra-lightweight distance-3 operators minimize term and stabilizer weight
def majorana_op(qubits, op_num):
    return QubitOperator(pauli_string_to_of(build_dist3_op_strings(qubits, op_num)))
\end{lstlisting}
\caption{The evolved constructor, verbatim: the complete mutable region of
the winning distance-5 program ($272\to168$ panel qubits), exactly as the
search wrote it (only the seed's fixed instruction banner removed). The
rule: build a near-complete simulation graph, delete a Hamiltonian cycle
drawn from the \emph{complement} of the interaction graph, keep the
identity assignment and the plain distance-3 operators.}
\label{fig:constructor-source}
\end{figure}

The reported rows combine three generated constructors and deterministic
post-search validation campaigns (Table~\ref{tab:program-lineage}). The search
services propose source code; distances, logical-failure rates, and transfer
claims are recomputed offline from the resulting constructors by the strict
verifier.

\begin{table*}[t]
\centering
\caption{Program lineage for the constructors used in the result tables. The
AlphaEvolve panel winner is the compact source in Fig.~\ref{fig:constructor-source};
the ShinkaEvolve replication starts from the same $272$-qubit seed but writes a
larger 2-factor dynamic program that is used for transfer beyond the original
panel. The floor-family constructor is a second ShinkaEvolve run against the
five-qubits-per-mode objective.}
\label{tab:program-lineage}
\scriptsize
\begin{ruledtabular}
\begin{tabular}{lll}
role & search source & use / validation \\
\colrule
dense panel & AlphaEvolve & H4/H6/LiH; exact distances $+$ $\pL$ \\
dense transfer & ShinkaEvolve & BeH2/H2O/NH3/CH4/N2; census, exact distances $+$ $\pL$ \\
floor family & ShinkaEvolve & $5n_{\rm modes}$ rows; CH4 fallback; exact distances $+$ $\pL$ \\
controls/baselines & deterministic + ShinkaEvolve & B2 random, single-stage, basis searches \\
\end{tabular}
\end{ruledtabular}
\end{table*}

\subsection{Resources and logical error at matched distance}
\label{sec:results-efficiency}
At exact distance 5 the evolved encodings use $4.2$--$5.0\times$ fewer
data qubits per molecule than the surface route and have
$3.4$--$8.2\times$ lower code-capacity logical-failure rates at
$p{=}10^{-3}$ under finite-weight minimum-weight decoder tables with
certified truncation brackets (Table~\ref{tab:headline}); the baseline is \emph{per-mode}
concatenation, one independent distance-5 block per JW qubit. The evolved
encodings use $48/60/60$ physical qubits on H4/H6/LiH, versus
$200/300/300$ for textbook JW followed by one $[[25,1,5]]$ surface patch
per mode ($\mathbf{4.2}$--$\mathbf{5.0\times}$ fewer; $4.8\times$ on the
panel sum, $168$ vs.\ $800$) and $136/204/204$ for the per-mode
$[[17,1,5]]$ color code ($2.8$--$3.4\times$); the panel is also leaner
than its seed ($272\to168$, $-38\%$). Two caveats temper the $38\%$:
the seed was a graduated prior winner that carried redundancy, so part of
the saving is undoing earlier over-provisioning; and the dominant mechanism
is the valence-reduction arithmetic on a still-near-complete graph, with
the interaction-aware choice of \emph{which} edges to drop primarily
protecting term weight. The rungs at other fixed distances frame the
compression (Table~\ref{tab:efficiency}; App.~\ref{app:rungs}).

Every distance in Table~\ref{tab:headline} is \emph{exact}: exhaustive
enumeration through weight 4 finds zero logical operators, and complete
weight-5 enumeration exhibits each code's minimum-weight logicals
($32$--$168$, led by weight-5 mode-parity and edge-product operators;
App.~\ref{app:semantics}, Table~\ref{tab:constructionspecs}). Applied unchanged beyond the panel, the
constructor's 2-factor generalization emits encodings for NH3 ($16$ modes),
CH4 ($18$), and N2 ($20$) at $112/144/180$ qubits ($3.6\times$/$3.1\times$/
$2.8\times$ fewer than their surface routes), with exact distances $5$,
$5$, and $6$: at $180$ qubits the complete weight-5 enumeration
($3.6\times10^{11}$ Paulis) finds \emph{zero} weight-5 logicals, and a
complete weight-6 enumeration exhibits the $210$
weight-6 ones (Table~\ref{tab:constructionspecs}). The shrinking ratio is
the dense family's quadratic qubit growth at work; the floor family of
\S\ref{sec:results-floor} removes it.

Raw data-qubit count at fixed $(k,d)$ is not by itself a sufficient resource
metric: generic constructions near the Gilbert--Varshamov bound give
$[[{\sim}35$--$45,12,5]]$ codes that beat every row of
Table~\ref{tab:headline} on raw count. The comparison here is narrower: it is a
code-capacity memory comparison against the surface route (per-mode JW $+$
$[[25,1,5]]$ surface), chosen because it has a direct logical-operation
interpretation. Generic
GV-bound codes and joint-block qLDPC memories encode JW qubits (the
bivariate-bicycle $[[72,12,6]]$ code~\cite{bravyi2024} reaches near
data-qubit parity, $72$ vs.\ $60$ at $12$ modes), so each Hamiltonian term
becomes a \emph{logical} Pauli on $O(n_{\rm modes})$ logical qubits (physical
support $O(n_{\rm modes}\,\dist)$ once implemented) via lattice surgery or
logical ancillas. The evolved encodings instead apply their encoded fermionic
terms natively as physical operators of measured bounded weight (maximum
$12$--$22$ across every code in this paper). We
therefore treat qLDPC/GV blocks as important adjacent baselines rather than as
resolved head-to-head competitors; a symmetric circuit-level analysis of
logical operations and syndrome extraction is future work
(\S\ref{sec:disc-limits}).

\begin{table*}[t]
\centering
\caption{The surface-route comparison at matched distance: each code (data qubits
$\nq$, exact distance by exhaustive enumeration) against the per-mode textbook
JW$+[[25,1,5]]$ surface route. The code $\pL$ intervals are analytic
expansions exact through weight $W{=}4$ with the full remaining tail carried
as a truncation bracket; the surface route uses the same decoder construction
with a weight-$6$ table for the surface block. Ratios are evaluated at
$p{=}10^{-3}$ and hold across the full truncation bracket: every code's
interval is disjoint from (and below) its surface route's at
$p\le 5\times10^{-3}$ (\S\ref{sec:results-pl}). Held-out
molecules were never scored during any search; the BeH2/H2O floor rows
are post-run applications of the floor rule, and all values are reproduced
from internal records by the strict verifier (App.~\ref{app:provenance}).}
\label{tab:headline}
\begin{ruledtabular}
\begin{tabular}{llccccc}
molecule & code & $\nq$ & $\dist$ (exact) & surf.\ route $\nq$ & code $\pL$ bracket & surf.\,route$/$code \\
\colrule
H4-chain & dense & \textbf{48} & 5 & 200 & $[2.27,2.29]\!\times\!10^{-7}$ & $3.6\times$ \\
H6-chain & dense & \textbf{60} & 5 & 300 & $[3.65,3.70]\!\times\!10^{-7}$ & $3.4\times$ \\
LiH      & dense & \textbf{60} & 5 & 300 & $[3.23,3.28]\!\times\!10^{-7}$ & $3.8\times$ \\
\colrule
BeH2 (held-out) & dense & 84 & 5 & 350 & $[1.82,2.11]\!\times\!10^{-7}$ & $7.9\times$ \\
H2O (held-out)  & dense & 84 & 5 & 350 & $[1.82,2.11]\!\times\!10^{-7}$ & $7.9\times$ \\
BeH2 (held-out) & floor & \textbf{70} & 5 & 350 & $[1.98,2.10]\!\times\!10^{-7}$ & $7.2\times$ \\
H2O (held-out)  & floor & \textbf{70} & 5 & 350 & $[1.98,2.10]\!\times\!10^{-7}$ & $7.2\times$ \\
\colrule
NH3 & floor & \textbf{80}  & 5 & 400 & $[2.00,2.22]\!\times\!10^{-7}$ & $8.2\times$ \\
N2  & floor & \textbf{100} & 5 & 500 & $[3.22,3.91]\!\times\!10^{-7}$ & $6.3\times$ \\
\end{tabular}
\end{ruledtabular}
\end{table*}

\subsection{A linear-scaling family at the qubit floor}
\label{sec:results-floor}
The family analysis of \S\ref{sec:results-family} exposes a structural
floor of five qubits per mode; a second search kept the strict distance-5
gate but rewarded qubit count against $Q_{B}=5\,n_{\rm modes}$, on a
training panel spanning the sizes where the dense rule's overhead grows
(H6, NH3, N2; $12/16/20$ modes), and found a constructor reaching that
floor on the tested sizes. The rule, again legible (verbatim core in
Fig.~\ref{fig:floor-source}, App.~\ref{app:rungs}): for
$n_{\rm modes}>12$, place the modes on a ring ordered by interaction
strength, take \Gsim{} as a degree-$9/10$ circulant $C_{n}(1,2,3,4,x)$
whose valence pins every vertex at $\lceil\deg/2\rceil=5$ qubits, then scan
the insertion order to bias the fundamental cycle basis toward light
stabilizers. No ancillas; the qubit count is $5\,n_{\rm modes}$ exactly,
and the per-mode-surface ratio becomes a constant $5.0\times$ in place of
the dense rule's shrinking ratio ($4$--$5\times$ at $8$--$12$ modes,
$2.8\times$ by $20$). The training codes at $12$, $16$, and $20$ modes
have exact distance 5
(Table~\ref{tab:constructionspecs}); at $12$ logical qubits the $60$-qubit
code undercuts the $[[72,12,6]]$ block's $72$ data qubits.

Applied unchanged to molecules never seen during the run, the rule
\emph{holds} on BeH2 and H2O ($14$ modes) at $70$ qubits each, with exact
distance 5 (beating the dense rule's $84$ on the paper's own held-out
panel), and \emph{fails} on CH4 ($18$ modes; certified $\dist\ge3$ only),
where the dense rule does hold at $144$ qubits. The practical deployment is
therefore a certified hybrid: try the floor family, certify; on failure
fall back to the dense rule, certify. The floor family also has measured
error rates: under the same finite-weight decoder tables its codes beat their surface routes by
$6.3$--$8.2\times$ at $p{=}10^{-3}$ (Table~\ref{tab:headline}), with
weight-3 failure fractions in line with the dense family's
(App.~\ref{app:pldata}); post-hoc term weights are within ${\sim}2\times$
of the dense rule's (maximum $12$--$22$). The qubit advantage over
per-mode concatenation is thus size-independent at the sizes tested where the
floor rule certifies. The verifier-grounded analysis of the first artifact
defined a new objective, and the search against that objective produced a
constructor at the predicted floor, under a hard \$$20$ budget
(\S\ref{sec:methods}).

\subsection{Logical error under finite-weight decoding tables}
\label{sec:results-pl}
\begin{figure}[t]
\centering
\includegraphics[width=\columnwidth]{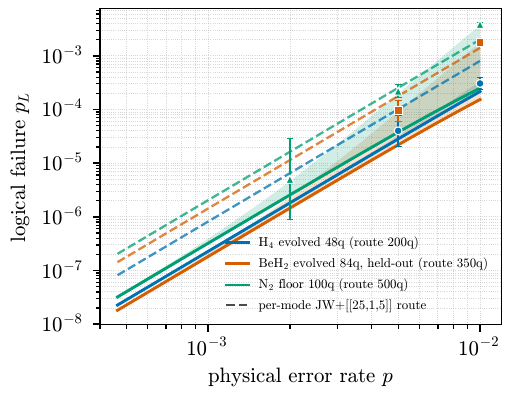}
\caption{%
  Logical failure $\pL(p)$ under finite-weight minimum-weight decoding tables,
  one representative code per tier (training / held-out / floor).
  Solid: analytic expansions from exhaustive per-weight failure counts,
  truncation bracket shaded. Dashed: the same molecule's surface route
  (JW$+[[25,1,5]]$, $1-(1-p_{\rm block})^{n_{\rm modes}}$, full
  weight-6 block table). Markers: $200$k-shot Monte-Carlo validation,
  Wilson $95\%$ CIs. All eleven codes and full grids:
  Table~\ref{tab:headline}, App.~\ref{app:pldata}.%
}
\label{fig:pl}
\end{figure}

With the finite-weight decoder tables of App.~\ref{app:eval}, every evolved and floor code has a lower logical-failure rate than
its surface route at $p{=}10^{-3}$, and the advantage grows with molecule size
(Fig.~\ref{fig:pl}). The protocol treats both sides identically: deterministic
minimum-weight decoding, with a full
syndrome table through weight 6 for the $[[25,1,5]]$ surface block (which
also pins the block's distance exactly at 5; App.~\ref{app:pldata}) and
breadth-first tables through weight 4 for the evolved and floor codes. Exhaustive classification of every weight-$\le4$
error gives exact failing counts ($f_1{=}f_2{=}0$ for every code; exact
$F_3$, $F_4$), which yield analytic $\pL(p)$ expansions with certified
brackets, validated by $200{,}000$-shot Monte-Carlo sampling with the same
tables: $29$ of $33$
cells land inside the bracket at $p\ge2\times10^{-3}$; the four outliers
are low-count statistics (App.~\ref{app:pldata}).
The numbers: on H4 at $p{=}10^{-3}$ the evolved $48$-qubit code has
$\pL\in[2.27,2.29]\times10^{-7}$ against the $200$-qubit surface route's $8.2\times10^{-7}$:
$3.6\times$ lower at $4.2\times$ fewer qubits. Across the panel the
advantage is $3.4$--$8.2\times$ and \emph{grows} with molecule size: the
surface route pays a linear union penalty ($n_{\rm modes}$ blocks are
$n_{\rm modes}$ chances to fail) while the joint code's weight-3 failure
count grows far slower than $\binom{n}{3}$---degeneracy corrects a growing
share of weight-3 errors, the failing fraction $f_3$ \emph{dropping} from
$1.31\times10^{-2}$ at $48$ qubits to $1.9\times10^{-3}$ at $84$. On the
held-out molecules the $84$-qubit codes have $7.9\times$ lower $\pL$ than
their surface route at $p{=}10^{-3}$. These numbers require a decoder table
deep enough to resolve the relevant errors: a shallow weight-2 table would
count every weight-$\ge3$ error as a failure, collapsing the measured $\pL$
on both routes to the ceiling $P(\mathrm{wt}\ge3\,|\,n)$ and inverting the
held-out verdict (App.~\ref{app:pldata}). The brackets
certify the win at $p\le5\times10^{-3}$ for every code; at $p{=}10^{-2}$
the $84$- and $100$-qubit comparisons are unresolved. Circuit-level
caveats, symmetric for both routes: \S\ref{sec:disc-limits}.

\subsection{The climb stage and what it teaches}
\label{sec:results-climb}
The same machinery under the distance-maximizing objective (the climb
stage) shows both the method's reach and its central failure mode: what is
retained is governed by the objective, not the search alone. Seeded with a
genuine distance-1 cell, the search climbs the strict-distance staircase to
$\dist{=}7$ ($\mc{=}3$) on all three molecules at $98$ qubits each ($684$
candidates hold strict $\dist{\ge}3$, $96$ hold strict $\dist{\ge}7$). But
the winning artifact is resource-blind, and because it is a program, that
structure is directly legible: the mutable body is three to six lines, a
dense \texttt{complete\_graph} on $\max(n_{\rm modes}{+}2,14)$ vertices,
the identity map, and a \emph{supplied} Clifford-rotated operator. The
distance is genuine (independently re-verified); the objective strongly
favors this resource-blind basin, because the qubit penalty is logarithmic
while a distance tier is worth a fixed bonus. The under-specified objective is visible in the
source code, a diagnostic unavailable when the optimized object is a
parameter vector; the population stays diverse to the end, continually
proposing qubit-lean structured graphs that the objective discards for not
reaching the dense graph's distance. The distance-3
control completes the picture: at a rung where the seed is already at its
qubit floor, the compression objective returns exactly zero reduction.

A matched-compute random-search baseline (B2) sharpens what the evolution
contributed. Drawing uniformly over the same three axes from a ten-family
graph menu \emph{that includes the dense complete-graph families the
evolution discovered}, random search reaches strict $\dist\ge5$ on $8.8\%$
of matched-budget samples ($132/1{,}500$; $57.7\%$ conditional on drawing
$K_m$, $25.9\%$ on $K_m$-minus-2-factor)---but \emph{never} from the
sparse families a priori intuition suggests ($0$ of $11{,}228$ draws over
grids, lattices, paths, and hyperbolic tilings). At $20{,}000$ samples it
rediscovers $60$-qubit $\dist{=}5$ codes on H6/LiH inside the known-good
$K_{12}$-minus-2-factor family, but never matches the evolved H4 economy
($50$ qubits minimum vs.\ $48$), and the interaction-aware
constructions (the complement-drawn deletion, the floor circulants) never
appear. Random search rediscovers instances within a known-good family;
identifying the families, the interaction-aware selection, and the
qubit-optimal padding is the constructor's content.

\subsection{Domain of validity of the discovered family}
\label{sec:results-family}
The discovered rule defines a code family whose domain of validity the
same verifier can chart---and the chart is subtler than a first sweep
suggested. A $35$-point sweep of the bare family (GSE on $K_N$ minus a
2-factor, \emph{block-canonical} vertex labeling) suggested a parity law
in $N$---strict $\dist\ge5$ failing for every deleted 2-factor at
$N\equiv2\ (\mathrm{mod}\ 4)$---while the constructor demonstrably
certifies real molecules at those sizes. An exhaustive $281$-point census
resolves the contradiction: \emph{over the tested census, certification varies
with the vertex labeling class of the deleted cycles, not just their cycle type}
(App.~\ref{app:family}). Under block-canonical labeling no type certifies
at $N\equiv2\ (\mathrm{mod}\ 4)$ ($0$ of all $124$ types at
$N\in\{10,14,18,22\}$); under spin-\emph{alternating} labeling (cycle
vertices alternating between the spin-$\alpha$ and spin-$\beta$ blocks)
\emph{every} all-even type certifies at $14\le N\le22$, including the
single Hamiltonian cycle; $N{=}10$ fails in every class; at odd
per-vertex qubit count ($N=12,16,20$) block labelings certify
type-dependently (all-even types always). The mechanism is chemistry: the
constructor's minimum-weight-2-factor program weights mode pairs by
co-occurrence in \texttt{fer\_op}, and in the blocked spin convention
cross-spin pairs are the cheap pairs, forcing even, strictly
spin-alternating deleted cycles---the certifying class ($6{+}4{+}4$ on
BeH2/H2O, $4{+}4{+}4{+}4$ on NH3, $6{+}6{+}6$ on CH4;
Table~\ref{tab:constructionspecs}); on uniform synthetic weights the
tie-breaks pick block-labeled cycles (the failing class), reproducing
its observed synthetic failures at $M\in\{14,15,18,22\}$
(App.~\ref{app:family}). The success on real chemistry at the bad-parity
sizes is therefore explained by chemistry-induced spin structure in the tested cases; the
earlier ``parity law'' survives only as its block-labeling restriction;
and the statement remains a characterization, not a proof; per-instance
verification stays mandatory.

The census also gives a structural bound: vertex operators are logicals of
weight $\lceil v/2\rceil$ on a valence-$v$ active vertex, so holding
$\dist{=}5$ requires active valence $\ge9$---a floor of ${\sim}5$ qubits
per mode that no constructor in this family can undercut, while the dense
rule pays ${\sim}N^{2}/2$. That floor became the objective of
\S\ref{sec:results-floor}; in the tested cases, the family certifies distance
5 (and distance 6 at the $180$-qubit point).

\section{Discussion}
\label{sec:discussion}

\subsection{Toward AI-assisted quantum design}
\label{sec:disc-outlook}
We read this case study as evidence for a design pattern, not for a claim
that language-model search ``understands'' quantum error correction. The
pattern: evolve a constructor, so discoveries are design rules testable
on held-out instances; channel all fitness through an exact, immutable
verifier; pose the objective as a resource trade at a held constraint,
so brute-force constructions are disfavored. None of this is specific to fermionic encodings: decoders,
syndrome-extraction circuits, Trotter ordering, compilation, and routing
share the shape. Two extensions would strengthen
the pattern. Pairing the search with a proof assistant could promote
per-instance certificates to family-wide
statements~\cite{lean4,dennis2002}; the labeling-class characterization of
\S\ref{sec:results-family} is a natural conjecture to formalize---and it
has already guided one follow-up search, with its floor becoming the objective
that produced the family of \S\ref{sec:results-floor}. Evaluator throughput
is also an important target, since it bounds the method's reach.

\subsection{Operating lessons}
\label{sec:disc-lessons}
Most of the human effort in this project did not go into designing
encodings directly. It went into running the search well: deciding what to
ask for, reading what came back, and changing the setup when the results
were unhelpful. A recent study of mathematicians working with an
evolutionary coding agent reaches a similar conclusion, arguing that these
systems are best treated as instruments to be operated rather than
assistants to be delegated to~\cite{bauerle2026}.

The most important lesson was also the most basic: the search can only
optimize what its edits actually move. Our first stabilizer-weight search
returned a null result that turned out to be an artifact of the harness
rather than a fact about the codes---the basis emission was fixed
\emph{outside} the editable block, so no edit the proposer could make
changed the quantity being rewarded (\S\ref{sec:disc-limits}). Before
reading anything into a stalled search, we learned to check that what the
proposer can edit is actually coupled to what the objective measures.
Evaluator speed proved nearly as decisive: exact verification is what makes
the scores worth trusting, but its cost is paid once per candidate and
thousands of times per run, and we rewrote several hot paths in the verifier,
capping every candidate at a fixed wall-clock limit (\S\ref{sec:methods}),
before iteration was cheap enough to be informative.

Context turned out to be a tunable input in its own right. What we told the
proposer---the problem semantics, the exact scoring formula, the available
library surface, the baseline numbers---visibly changed the proposals we got
back during warmup, and we came to treat the context pack as a versioned part
of each run rather than as boilerplate. A final lesson concerned staging the
objective when one reward must give way to another. A matched-budget control
optimized a \emph{single-stage} objective, shaping the distance reward and
merely \emph{unlocking} the compression term once the target distance was
reached, starting from the distance-1 seed; it cleared the distance-5 gate
within ${\sim}10\%$ of its budget and then compressed almost nothing over the
remaining $90\%$, ending in a dense ${\sim}291$-qubit basin, whereas the
staged pipeline reaches $168$ qubits at the same spend. Once at the gate, the
compression gradient is too weak to pull the search off the worst passing
code (\S\ref{sec:results-climb}), so separating the two stages lets the second
devote its full budget to compression.

\subsection{Limitations}
\label{sec:disc-limits}
Exact-distance certificates and a code-capacity error advantage are
necessary but not sufficient for a hardware advantage, and several gaps
remain. The model is code-capacity only: every certificate and every
$\pL(p)$ assumes i.i.d.\ depolarizing noise on the data qubits with one round
of perfect syndrome extraction. The encodings apply their fermionic terms as
bounded-weight physical operators, but that is a structural property, not a
fault-tolerance guarantee, and the circuit-level effective distance is
unmeasured for both routes; a circuit-level analysis---and its interaction
with complementary, hardware-level error-suppression methods such as
mid-circuit-measurement Clifford-noise reduction~\cite{ionq2026}---is future
work. The advantage is also specifically about \emph{memory}: the
$3.4$--$8.2\times$ improvement of \S\ref{sec:results-pl} bounds
stored-information error, not the error of executing dynamics, whose
${\sim}n^{3}$--$n^{4}$ operation count is the same for both routes. Within
that memory comparison the advantage grows with size, as degeneracy
suppresses weight-3 failures, but whether it persists or crosses over beyond
the tested sizes is open.

A second group of caveats concerns cost and optimality. The codes' stabilizer
generators are heavy---weight $11$--$16$ against the surface code's $4$---so
weight-$\ge9$ checks incur the cat-state or flag overheads that weight-4
checks avoid (App.~\ref{app:eval}). Re-choosing the generating set of a fixed
code cuts total generator weight by $29$--$38\%$, and a joint search over
graph and basis reaches maximum weight $10$ on H6 and LiH at the same distance
and qubit count; but H4 retains a maximum of $11$ that we certify optimal for
its stabilizer group, and reducing it further would require changing the code
or Hastings-style weight reduction~\cite{sabo2024}, a complementary lever our
fixed-qubit searches do not explore. The compression is also partly
seed-relative: part of the qubit saving is measured against the seed
construction (\S\ref{sec:results-efficiency}), making it an improvement over
that baseline rather than a proven optimum.

Finally, our central structural claim is a characterization, not a proof. The
labeling-class census of \S\ref{sec:results-family} is exhaustive over its
$281$ points but empirical, with no analytic proof of the alternation
boundary, and the constructor fails on CH4 where the dense rule succeeds;
deployment is therefore the certified hybrid, with per-instance verification
in either case. The molecular panel is likewise limited in scope, using the
minimal sto-3g basis, with larger bases untested.

\section{Related work}
\label{sec:related}

\emph{Distance-bearing fermionic encodings.} The closest prior art is the
Chen--Gorshkov--Xu (CGX) construction~\cite{chen2022}, to our knowledge
the only published intrinsic (non-concatenated) distance-5 fermionic
code. It is engineered for a 2D
square lattice of geometrically local fermions, with no native form for
dense molecular interaction graphs; a machine-checked certificate on
Clifford-plus-permutation invariants confirms the evolved code is
non-isomorphic to it on every molecule (App.~\ref{app:provenance}). The
original superfast/GSE result~\cite{setia2018} proves distance 3
analytically at interaction-graph vertex degree ${\ge}6$
(App.~\ref{app:semantics});
square-lattice codes, compact mappings, and Majorana-loop stabilizer
codes~\cite{steudtner2019,derby2021,jiang2019} develop the lineage on
structured geometries; our group's prior GSE-for-molecules
work~\cite{brown2025} demonstrates error \emph{detection} (distance 2) on
molecular interaction graphs, alongside an analytic distance-$(2N{+}1)$
family with constant weight-6 stabilizers on \emph{linear} interaction
graphs. This work adds the first decoder-verified distance-5 (and one distance-6)
GSE encodings on dense molecular graphs---beyond the distance-3 ceiling of
prior superfast/GSE constructions---through program search rather than hand
construction. Lattice-oriented approaches reach high distance by other means: searched
intrinsic encodings on Fermi--Hubbard geometries~\cite{simkovic2024}, and
concatenated constructions in surface and color codes that attain arbitrary
distance at constant stabilizer weight~\cite{algaba2025,landahl2021}---most
directly Wei et al.~\cite{wei2025}, who concatenate a small-distance
fermion-to-qubit code with a high-distance fermionic color code in 2D and
3D. These target lattice fermions rather than dense molecular Hamiltonians,
and inherit their distance from an outer code at concatenation overhead;
our distance-5 codes are intrinsic, at the cost of higher stabilizer weight
(\S\ref{sec:disc-limits}). Joint-block qLDPC memories (such as the bivariate-bicycle
$[[72,12,6]]$ code of Bravyi et al.~\cite{bravyi2024}) encode many JW
qubits per block and reach near data-qubit parity with the evolved
encoding, at the cost of high-weight logical operators for the molecular
Hamiltonian terms (\S\ref{sec:results-efficiency}). We therefore frame the
headline comparison as route-specific: against independent per-mode
concatenation in a code-capacity memory model, not as a dominance claim over
joint-block qLDPC memories or other generic code families whose logical
operation costs require a separate circuit-level analysis.

\emph{Search-discovered codes and objective design.} Search-discovered
fermion-to-qubit \emph{mappings} exist (simulated annealing~\cite{yu2025},
enumeration~\cite{chiew2021}) but optimize Pauli weight at distance-1.
\v{S}imkovic et al.~\cite{simkovic2024} already search distance-bearing
fermionic encodings, via a fixed-form enumeration over
translationally-invariant local encodings on the square lattice rather
than program synthesis over constructors, so we make no first-to-search
claim. The program-search lineage runs from
FunSearch~\cite{romeraparedes2024} to AlphaEvolve~\cite{novikov2025}, our
discovery engine. While preparing this manuscript we became aware of two concurrent efforts
applying LLM-guided evolutionary search to quantum error correction:
Cruz-Benito et al.~\cite{cruzbenito2026} evolve code-\emph{generating}
programs for bivariate-bicycle qLDPC ans\"atze with staged distance
certification, and Liu and Marquardt~\cite{liu2026} pair an LLM with a
structured algebraic mutation grammar to evolve lifted-product qLDPC
families. LLM-assisted lifted-product search~\cite{cain2026} and
distance-certification benchmarks~\cite{webster2026} are further surrounding
context. The efforts are
complementary: they search a parameterized qLDPC catalogue for strong
general-purpose memories, we evolve instance-conditioned constructors for
fermionic encodings; on data qubits their $[[72,12,6]]$-class blocks and
our $60$--$100$-qubit codes are close, and the differentiating cost is
logical-operation weight (\S\ref{sec:results-efficiency}). Their
evaluator-hygiene lessons are measurement-side (staged certification
inside the loop); ours objective-side (climb-then-compress). Reward
hacking and specification gaming are well documented for
control policies and learned value
functions~\cite{amodei2016,skalse2022,pan2022,krakovna2020}; here the
under-specified objective is visible in source code on a real design task, and
the over-provisioned-seed-plus-compression-reward recipe turns the distance
reward into a held constraint.

\section{Conclusion}
\label{sec:conclusion}

This case study demonstrates that evolutionary program synthesis with an exact
verifier in the loop can be used as a design tool for quantum computing. The
search discovered fermionic-encoding constructors whose codes have \emph{exact}
code distance 5 (and 6 on one 20-mode instance) on the molecular instances
tested---to our knowledge the first GSE/superfast encodings beyond distance 3
for dense molecular Hamiltonians. As secondary resource descriptors, these
codes use $4.2$--$5.0\times$ fewer data qubits than the textbook surface route
(per-mode JW $+$ $[[25,1,5]]$ surface) and, in a code-capacity memory
comparison at $p{=}10^{-3}$ under finite-weight minimum-weight decoder tables
with truncation brackets, have $3.4$--$8.2\times$ lower logical-failure rates
on the tested training and held-out molecules. The dense constructor's domain of validity
was mapped to an empirical labeling-class characterization, and that chart's
qubit floor became the objective of a second search whose circulant family
holds exact distance 5 at the tested $12$-, $14$-, $16$-, and $20$-mode
instances, with an $18$-mode CH4 failure handled by certified dense-rule
fallback. A key variable was the objective: rewarding distance directly
selected a resource-blind dense-graph basin at genuine distance 7, whereas
holding distance fixed and rewarding compression from the climb's verified
seed selected structured rules.

This does not show that the codes exceed what careful human design could
achieve, nor does it establish a circuit-level fault-tolerance advantage. The
comparison is a data-qubit, code-capacity memory comparison against the scoped
surface route, and the evolved codes carry heavier stabilizer checks than the
surface code's weight-4 ones. We read the result as a step from \emph{selecting} codes out of a
fixed catalogue toward \emph{synthesizing} them against an exact verifier, with
the synthesis only as good as its objective. Future work will address
the limitations in \S\ref{sec:disc-limits}, starting with circuit-level
fault-tolerance analysis, while also exploring richer design objectives aside from pure distance and resource efficiency.

\begin{acknowledgments}
We thank Federico Rodriguez, Christopher Penny, Clara Buenker, Adrian
Jones, Anant Nawalgaria, Skander Hannachi, and Vishal Agarwal, along with
the rest of the AlphaEvolve and AI for Science teams at Google Cloud, for
access to the AlphaEvolve Early Access Program and for their support and
technical guidance throughout this work.
\end{acknowledgments}

\appendix

\section{Distance semantics and certification}
\label{app:semantics}

\paragraph{Canonical semantics.}
\label{par:semantics-canonical}
All distance and logical-failure statements in this paper use one
equivalence relation: two Pauli errors are \emph{equivalent} if and only if
they differ by an element of the stabilizer group
$\langle\mathrm{stab\_gens}\rangle$ generated by the closed-loop operators
alone. The mode-parity vertex operators $V_i$ are encoded \emph{logical
observables}. An error equal to a product
of $V_i$ therefore changes the encoded state and counts as a logical
failure. Under this relation, a distance certificate enumerates all Pauli
errors up to the target weight and demands that \emph{inequivalent} errors
produce distinct syndromes; a syndrome shared by two inequivalent errors is
counted as a failure even when a maximum-likelihood decoder could exploit
degeneracy, so every certified $\mc$ (and hence every certificate-derived
$\dist\ge2\,\mc+1$) is a conservative lower bound. Decoding succeeds exactly
when the residual operator---the sampled error times the applied
correction---lies in $\langle\mathrm{stab\_gens}\rangle$; a residual in
$\langle V_i\rangle$ is a logical failure. The same relation, applied by
the same verifier, underlies the distance certificates, the exact-distance
enumerations, and the $\pL(p)$ measurements. The permissive convention
$\langle\mathrm{stab\_gens}\cup V_i\rangle$ reports strictly larger
``distances'' and is never used here; the strict choice is the faithful
numerical counterpart of the founding GSE construction, in which the
stabilizer group is generated by the closed-loop operators alone and the
$V_i$ are the encoded fermionic logical operators~\cite{setia2018}.

\paragraph{Exact distances.}
Certificates give lower bounds; the distances quoted in this paper are
\emph{exact}, established by exhaustive enumeration. For each code we
enumerate every Pauli operator of weight $\le5$ that commutes with all
stabilizer generators, and test GF(2) membership in
$\langle\mathrm{stab\_gens}\rangle$: weights $1$--$4$ contain zero
logical operators for every code (re-confirming the strict $\mc{=}2$
certificates), and the weight-5 stratum exhibits each code's minimum-weight
logicals---$32$--$168$ distinct weight-5 logical operators per code, led by
weight-5 mode-parity (vertex) operators and edge-product strings
(Table~\ref{tab:constructionspecs}). For the $180$-qubit dense-N2 code the
complete weight-5 enumeration ($3.6\times10^{11}$ Paulis) finds zero
logicals, and a complete weight-6 enumeration---meet-in-the-middle, pairing
weight-3 halves by matching syndrome so the weight-6 search costs
$O(n^3)$ rather than $O(n^6)$---exhibits $210$ weight-6 logicals, pinning
$\dist=6$ exactly. Every witness passed an
independent re-verification gate (commutation against every generator plus
GF(2) non-membership) before being recorded.

\paragraph{Analytic vs.\ numerical certification.}
It is worth being precise about how the prior distance-3 result was
established, because our methodology differs and the difference is what
makes a distance-5 claim tractable. Setia et al.~\cite{setia2018} prove
distance-3 \emph{analytically}: their Theorem~1 shows, by a 3-connectivity
argument tied to vertex degree $\ge6$, that no logical operator of weight
$<3$ can exist for the GSE on a qualifying graph---a once-and-for-all
statement about a graph family that invokes no decoder and no enumeration,
with correction of more than one error left open. By contrast, our
distances are \emph{measured per instance}: exhaustive enumeration on each
concrete code. The approaches are complementary---an analytic sufficiency
theorem at $\dist{=}3$ versus an exact numerical verifier that reaches
exact $\dist{=}5$--$6$ on concrete molecules; promoting per-instance
certificates to family-wide statements is discussed in
\S\ref{sec:disc-outlook}.

\section{The evolutionary search apparatus}
\label{app:apparatus}

This appendix specifies the objective the search optimized and the
anti-tampering defenses on the evaluator.

\subsection{Objective functions}
\label{app:objectives}

Both regimes are additive, resource-normalized, per-molecule scores aggregated
across the panel by a consistency-rewarding harmonic mean with offset, and
share an identical resource term: relative to the conventional
$\mathrm{JW}+[[5,1,3]]$ QEC reference for each molecule, a candidate is
penalized by its excess physical qubits, Hamiltonian term weight, and
stabilizer weight,
\begin{equation}
\label{eq:resource}
R(m) \;=\; - \lambda_q \log_{10}\!\tfrac{\nq}{Q_{\rm base}}
            - \lambda_w \log_{10}\!\tfrac{W}{W_{\rm base}}
            - \lambda_s \log_{10}\!\tfrac{S}{S_{\rm base}},
\end{equation}
with $\lambda_q{=}200$, $\lambda_w{=}\lambda_s{=}20$ per decade. The
log-ratio, additive (never multiplicative) form follows the standard design
principle that an objective should be a monotone weighted sum of normalized
terms, which avoids the pathological gradients that reward hacking
exploits~\cite{skalse2022,pan2022}.

\paragraph{Regime~A (distance as reward; the climb stage).}
Each molecule contributes a smooth distance ladder on top of the resource
term,
\begin{equation}
\label{eq:regimeA}
\mathrm{raw}_A(m) \;=\; \sum_{k\ge 1}\beta_k\, d_k(m)\;+\;R(m),
\qquad \beta_k = 40\cdot 2^{\,k-1},
\end{equation}
where $d_k(m)\in[0,1]$ is the strict fraction of weight-exactly-$k$ Pauli
errors that the code corrects, computed lazily up the ladder. The geometric
weights $\beta_k=40,80,160,320,640$ make saturating a new distance tier
($d_k{:}\,0{\to}1$) dominate any partial progress on lower tiers---encoding
``climb distance first''---while the always-on resource term $R(m)$ provides
a secondary pull toward leaner codes within each tier.

\paragraph{Regime~B (distance as a held constraint; the compression stage).}
A hard gate requires the candidate to hold the target strict distance on
\emph{every} molecule ($d_1{=}\dots{=}d_t{=}1$); a candidate that drops below
it receives a heavy penalty. Above the gate the score is a fixed hold-bonus
plus the resource term,
\begin{equation}
\label{eq:regimeB}
\mathrm{raw}_B(m) \;=\;
\begin{cases}
\textsc{hold\_bonus} + R(m), & d_1{=}\dots{=}d_t{=}1,\\[2pt]
\text{penalty}, & \text{otherwise.}
\end{cases}
\end{equation}
There is no reward for exceeding the target distance; the entire gradient
above the gate is the resource term.

\subsection{Anti-tampering defenses}
\label{app:evaluator}

The anti-tampering defenses referenced in the Method section:
candidates are AST-scanned for forbidden imports and calls before
execution; distance is never trusted from the candidate but recomputed by
the exact verifier; evaluation is deterministic; the scoring logic lives in
a file the model never sees; and the held-out panel is never exposed.
Every number in this paper is recomputed offline from logged run artifacts
by the strict verifier (App.~\ref{app:provenance}).

\section{Code evaluation and performance data}
\label{app:perf}

This appendix details the evaluation methodology and reports the resulting
construction specifications and logical-failure data.

\subsection{Evaluation methodology}
\label{app:eval}

\paragraph{Molecular panel.}
The training panel comprises three molecular Hamiltonians at the sto-3g
basis: the H4 hydrogen chain (8 spin-orbitals), the H6 hydrogen chain (12),
and LiH (12). Two molecules withheld entirely from the search form the
held-out panel: BeH2 (14) and H2O (14, a bent out-of-distribution geometry),
evaluated only after the search completes, on the top-ranked programs. The
panel deliberately mixes a size sweep (8 vs.\ 12 modes in training, 14 in
held-out) with a chemistry sweep (homonuclear chains, the ionic LiH, the bent
H2O), so that a constructor which merely overfits one mode count or one
interaction pattern is exposed. One consequence for the resource tables: H6
and LiH have the same mode count and, at sto-3g, near-identical
interaction-graph structure, so the structural metrics (qubit count, $\mc$)
do not separate them; for the qubit claim the effective panel is two sizes
($8$ and $12$ modes), and the chemistry sweep stresses the constructor's
structural heuristics rather than the headline qubit ratio.

\paragraph{Distance verification.}
$\mc$ is the largest weight $t$ for which the strict-coset verifier finds the
error-to-syndrome map injective up to the stabilizer group
(App.~\ref{app:semantics}). This is an exact, deterministic computation, so the
distance numbers carry no statistical uncertainty; the exhaustive
minimum-weight-logical enumerations that upgrade the resulting lower bounds
to exact distances are described in the same appendix.

\paragraph{Logical-failure measurement.}
All $\pL$ numbers use the same deterministic minimum-weight decoder
construction, applied identically to both routes. A syndrome-to-correction table is built
by breadth-first enumeration of \emph{all} Pauli errors in weight order
($0,1,\dots,W$; first writer to a syndrome wins, so every table entry is a
minimum-weight representative), with $W{=}4$ for the evolved and floor
codes and $W{=}6$ for the $[[25,1,5]]$ surface block. The depths are
asymmetric by necessity: the $25$-qubit, single-logical surface block is small
enough to enumerate to $W{=}6$ (which additionally pins its distance exactly at
$5$), whereas for the $48$--$100$-qubit evolved and floor codes exhaustive
weight-$W$ enumeration costs $\binom{n}{W}3^{W}$ and $W{=}4$ is the deepest
table common to all of them. The asymmetry is conservative for the comparison:
the evolved codes' unmodeled weight-$\ge5$ tail is carried explicitly in the
truncation bracket rather than assumed to decode, and because $f_1{=}f_2{=}0$
with every code at distance $\ge5$, the leading failure order is weight
$3$---well inside $W{=}4$. Decoding an error $E$
applies the table entry for its syndrome; \emph{success} iff the residual
(error times correction) lies in $\langle\mathrm{stab\_gens}\rangle$
(strict semantics, App.~\ref{app:semantics}). Exhaustive classification of
every weight-$\le W$ error under this decoder gives exact per-weight
failing counts $F_w$, from which
$\pL(p)=\sum_w F_w\,(p/3)^w(1-p)^{\,n-w}$ exactly through weight $W$; the
truncation tail $P(\mathrm{wt}>W)$ is carried as an explicit
lower/upper bracket. Monte-Carlo validation samples iid depolarizing noise
($200{,}000$ shots per point, Wilson $95\%$ CIs) through the same tables.
Thus the decoder table is exact through the stated weight cutoff, while the
reported $\pL(p)$ values are bracketed finite-weight series, not all-weight
closed-form logical-failure probabilities.
The route-level failure event is the same on both sides---the encoded
molecular state is corrupted---computed for the surface route as
$1-(1-p_{\rm block})^{n_{\rm modes}}$ over independent blocks. Syndrome
extraction is treated as perfect: the standard code-capacity setting,
appropriate for comparing routes, not for extracting a threshold or an
operational per-round error rate.

\paragraph{Baselines.}
The conventional QEC baseline is \emph{textbook} Jordan--Wigner---exactly one
logical qubit per spin-orbital---concatenated with an independent block per
logical qubit, the molecule failing if any block fails. We are careful to use
the textbook $n_{\rm modes}$ count: embedding JW-style operators in the GSE
machinery would allocate $\lceil\deg/2\rceil$ qubits per mode and overstate
the baseline $2$--$4\times$. The scoring objective normalizes against the
distance-3 member, $\mathrm{JW}+[[5,1,3]]$ (App.~\ref{app:objectives}); the
flagship resource comparison uses the distance-matched $[[25,1,5]]$ rotated
surface code (verified CSS-commuting with $24$ independent checks and exact
$\mc{=}2$ before use; its full weight-6 decode table additionally pins its
distance exactly at 5 via $160$ weight-5 logicals) and the $[[17,1,5]]$
color code. The joint-block comparison of \S\ref{sec:results-efficiency}
counts the bivariate-bicycle $[[72,12,6]]$ code~\cite{bravyi2024}: one block
per molecule covers up to $12$ JW logical qubits in $72$ data qubits at
distance $6$ with weight-6 checks; we do not simulate its $\pL$ or its
logical operations, so that comparison is on data-qubit count and the
structural cost of applying Hamiltonian terms only. On syndrome-extraction
cost, a simple data-plus-ancilla count under bare one-ancilla-per-check
extraction: the surface route costs $25+24=49$ per mode ($1568$
over the training panel); the evolved code costs $\nq$ data plus
$\nq-n_{\rm modes}$ checks, i.e.\ $2\nq-n_{\rm modes}$ ($88/108/108$ on
H4/H6/LiH, $304$ panel)---a ${\sim}5.2\times$ ratio, comparable to
the data-only $4.8\times$ since both sides roughly double. This omits the
cat-state or flag overhead that faithfully measuring weight-$\ge9$ checks
requires and weight-4 checks do not (\S\ref{sec:disc-limits}). The
hand-designed baseline (B1) sweeps Jordan--Wigner, Bravyi--Kitaev
superfast, and GSE on grid, triangular, hyperbolic, and hexagonal
substrates; every such family tops out at strict $\dist{=}3$ on the held-out
molecules (BeH2, H2O), where the evolved constructor---never tuned on
them---certifies exact $\dist{=}5$ (\S\ref{sec:results-efficiency}). The
random baseline (B2) is described in \S\ref{sec:results-climb}.

\subsection{Fixed-distance rungs and construction specifications}
\label{app:rungs}

Table~\ref{tab:efficiency} reports the fixed-distance compression apparatus
(the compression stage) across both distance rungs.
Table~\ref{tab:constructionspecs} gives the construction specification and
exact distance of every flagship encoding, so a reader can rebuild each
code without running the programs; Fig.~\ref{fig:floor-source} reproduces
the floor-family rule verbatim. Table~\ref{tab:floorfamily} compares the
floor and dense families per molecule.

\begin{table}[h]
\centering
\caption{The fixed-distance optimizer across rungs. Seed versus best holder
(panel-total qubits), the compression, and the fraction of candidates holding
the target distance panel-wide.}
\label{tab:efficiency}
\begin{ruledtabular}
\begin{tabular}{ccccc}
target $\dist$ & seed $\nq$ & best $\nq$ & compression & hold rate \\
\colrule
3 & 96  & 96  & $0\%$           & $446/1514$ ($29\%$) \\
5 & 272 & 168 & $\mathbf{-38\%}$ & $257/1411$ ($18\%$) \\
\end{tabular}
\end{ruledtabular}
\end{table}

\begin{table*}[t]
\centering
\caption{Construction specifications and exact distances for every flagship
encoding. ``Dense'' is the evolved constructor of
Fig.~\ref{fig:constructor-source} and its 2-factor generalization (a
near-complete graph minus a deleted edge set drawn from the complement of
$\Gint$); ``floor'' is the circulant rule of Fig.~\ref{fig:floor-source}
($5\,n_{\rm modes}$ qubits exactly; the fifth circulant offset $x$ is
selected per molecule by the rule's cycle-basis proxy). All codes use the
plain distance-3 operator basis. Distances are exact: exhaustive
enumeration finds zero logicals below $\dist$ and the listed number of
distinct minimum-weight logicals at $\dist$ (App.~\ref{app:semantics}).
Floor-H6 coincides with dense-H6 (the floor rule retains the dense rule at
$12$ modes).}
\label{tab:constructionspecs}
\scriptsize
\begin{ruledtabular}
\begin{tabular}{llcclcc}
family & molecule & modes & $\nq$ & construction & $\dist$ (exact) & min-wt logicals \\
\colrule
dense & H4-chain & 8  & 48  & $K_{10}$ ($8$ active $+$ $2$ ancilla spokes), ancilla--ancilla edge pruned & $5$ & 112 \\
dense & H6-chain & 12 & 60  & $K_{12}$ minus Ham.\ cycle from complement of $\Gint$ (Warnsdorff) & $5$ & 168 \\
dense & LiH      & 12 & 60  & $K_{12}$ minus Ham.\ cycle from complement of $\Gint$ (Warnsdorff) & $5$ & 144 \\
dense & BeH2     & 14 & 84  & $K_{14}$ minus spin-alternating 2-factor, type $6{+}4{+}4$ (DP) & $5$ & 88 \\
dense & H2O      & 14 & 84  & $K_{14}$ minus spin-alternating 2-factor, type $6{+}4{+}4$ (DP) & $5$ & 88 \\
dense & NH3      & 16 & 112 & $K_{16}$ minus spin-alternating 2-factor, type $4{+}4{+}4{+}4$ (DP) & $5$ & 56 \\
dense & CH4      & 18 & 144 & $K_{18}$ minus spin-alternating 2-factor, type $6{+}6{+}6$ (DP) & $5$ & 78 \\
dense & N2       & 20 & 180 & $K_{20}$ minus single Ham.\ cycle $[20]$ (documented DP fallback at $20$ modes) & $6$ & 210 \\
\colrule
floor & H6-chain & 12 & 60  & dense rule retained (already on the floor at $12$ modes) & $5$ & 168 \\
floor & BeH2     & 14 & 70  & $C_{14}(1,2,3,4,x)$ circulant, interaction-ordered ring & $5$ & 34 \\
floor & H2O      & 14 & 70  & $C_{14}(1,2,3,4,x)$ circulant, interaction-ordered ring & $5$ & 34 \\
floor & NH3      & 16 & 80  & $C_{16}(1,2,3,4,x)$ circulant, interaction-ordered ring & $5$ & 32 \\
floor & CH4      & 18 & (90) & $C_{18}$ degree-$9$ diametric variant & fails ($\ge3$ only) & --- \\
floor & N2       & 20 & 100 & $C_{20}(1,2,3,4,x)$ circulant, interaction-ordered ring & $5$ & 102 \\
\end{tabular}
\end{ruledtabular}
\end{table*}

\begin{figure*}[t]
\begin{lstlisting}[language=Python, basicstyle=\scriptsize\ttfamily,
    breaklines=true, breakindent=1.5em, columns=fullflexible,
    commentstyle=\itshape, showstringspaces=false]
# Hybrid floor-achieving family:
#
# * n = 12: use the best-performing dense-floor family, K_n minus a
#   carefully chosen 2-factor, then shape insertion/root order to bias
#   NetworkX's fundamental cycle basis toward lighter stabilizers.
#
# * n > 12: switch to a constant-degree active-only sparse family at the
#   structural 5-qubits-per-mode floor.  A small bank of degree-10
#   circulants is scored with a cheap cycle/path proxy after an
#   interaction-aware ring placement of modes.
#
# This preserves strict d_1=d_2=1 across the panel while keeping qubits
# linear in the number of modes.

# [... omitted: _pair_weights (interaction weights from fer_op.terms),
#      _interaction_ring_order (greedy ring placement of modes by
#      interaction weight), _cycle_basis_proxy, and the dense n<=12
#      branches carried over from the prior winner ...]

def _sparse_floor_graph(num_modes, weights, mode_order):
    """Select a 5-qpm active-only sparse family with basis-order shaping."""
    half = num_modes // 2
    candidate_specs = []

    def _push_offsets(raw_offsets):
        # [... omitted: normalize offsets mod n, deduplicate ...]
        degree = 0
        for s in offs:
            if num_modes % 2 == 0 and 2 * s == num_modes:
                degree += 1
            else:
                degree += 2

        # Keep only families that stay at the 5-qpm floor:
        # valence 9 or 10 both map to ceil(v/2)=5.
        if degree not in (9, 10):
            return

        spec = (offs, degree)
        if spec not in candidate_specs:
            candidate_specs.append(spec)

    # Proven degree-10 backbone family.
    _push_offsets((1, 2, 3, 4, 5))
    for extra in range(6, half):
        _push_offsets((1, 2, 3, 4, extra))

    if half - 1 > 5:
        _push_offsets((1, 2, 3, 5, half - 1))
        _push_offsets((1, 2, 4, 5, half - 1))

    # Conservative degree-9 family for larger even instances:
    # replace one paired chord with the diametric matching offset n/2.
    # This keeps qpm=5 while reducing cycle rank and potentially tsw.
    if num_modes % 2 == 0 and num_modes >= 18:
        _push_offsets((1, 2, 3, 4, half))
        if half - 1 > 4:
            _push_offsets((1, 2, 3, half - 1, half))
            _push_offsets((1, 2, 4, half - 1, half))
        if half - 2 > 4:
            _push_offsets((1, 3, 4, half - 2, half))
            _push_offsets((2, 3, 4, half - 2, half))

    # [... omitted: for each candidate circulant, scan root vertex,
    #      direction, and edge-insertion order; rank by the fundamental
    #      cycle-basis proxy, edge count, and the interaction-weighted
    #      path-length proxy; return the best graph ...]

if n_modes > 12:
    weights = _pair_weights(n_modes)
    mode_order = _interaction_ring_order(n_modes, weights)
    G_sim = _sparse_floor_graph(n_modes, weights, mode_order)
    mode_vertex_map = {mode: pos for pos, mode in enumerate(mode_order)}
\end{lstlisting}
\caption{The floor-family rule, in the search's own words. Core of
the winning floor-family program (ShinkaEvolve, generation $23$), verbatim
including comments; omitted helper bodies are marked. The rule: order the
modes on a ring by interaction strength, restrict to circulant graphs
$C_{n}(1,2,3,4,x)$ (degree-$9$ diametric variants for even $n\ge18$) whose
valence $9$--$10$ pins every vertex at the five-qubit floor
$\lceil v/2\rceil=5$, then pick the candidate whose insertion order yields
the lightest fundamental cycle basis.}
\label{fig:floor-source}
\end{figure*}

\begin{table}[t]
\centering
\caption{The floor family against the dense rule and the surface route,
per molecule. Training molecules above the rule; below it, molecules never
seen during the floor run. All ``holds'' rows are exact $\dist{=}5$
(Table~\ref{tab:constructionspecs}); CH4 fails the weight-2 tier and is
certified $\dist\ge3$ only (the dense rule holds there).}
\label{tab:floorfamily}
\begin{ruledtabular}
\begin{tabular}{lccccl}
molecule & modes & floor & dense & JW+surf.\ & floor status \\
         &       & $\nq$ & $\nq$ & {\footnotesize$[[25,1,5]]$} &  \\
\colrule
H6-chain & 12 & \textbf{60}  & 60  & 300 & $\dist{=}5$ exact \\
NH3      & 16 & \textbf{80}  & 112 & 400 & $\dist{=}5$ exact \\
N2       & 20 & \textbf{100} & 180 & 500 & $\dist{=}5$ exact \\
\colrule
BeH2     & 14 & \textbf{70}  & 84  & 350 & $\dist{=}5$ exact \\
H2O      & 14 & \textbf{70}  & 84  & 350 & $\dist{=}5$ exact \\
CH4      & 18 & 90           & \textbf{144} & 450 & fails ($\ge3$) \\
\end{tabular}
\end{ruledtabular}
\end{table}

\subsection{Full logical-failure data}
\label{app:pldata}

Table~\ref{tab:fw} reports the exact per-weight failure profile behind
Fig.~\ref{fig:pl} and the $\pL$ ratios of Table~\ref{tab:headline}:
exhaustive classification of every weight-$\le W$ Pauli error under the
finite-weight minimum-weight decoder tables (App.~\ref{app:eval}).
$f_1=f_2=0$ \emph{exactly} for every code, consistent with the strict
$\mc{=}2$ certificates. Two structural facts carry the head-to-head. The
dense evolved codes get \emph{better} with size---$f_3$ drops from
$1.31\times10^{-2}$ ($48$q) to $1.90\times10^{-3}$ ($84$q)---because more
stabilizers mean more degeneracy, so a larger share of weight-3 errors
decodes correctly; the surface route, by contrast, multiplies a fixed block failure
by $n_{\rm modes}$. And the floor codes behave like the dense family
($f_3=2.0$--$3.6\times10^{-3}$ at $70$--$100$q). Code identities visible in
the table: floor-H6 is byte-identical to evolved-H6, and the BeH2/H2O
pairs are isomorphic with identical counts. The analytic expansions are
exact through $W$ with the truncation tail carried as a bracket; at
$p{=}10^{-3}$ the brackets are $\le16\%$ wide, and at $p\le5\times10^{-3}$
every code's bracket is disjoint from (below) its surface route's, certifying the
ratios of Table~\ref{tab:headline}; at $p{=}10^{-2}$ the $84$q and $100$q
brackets overlap their surface routes and those cells are unresolved. Monte-Carlo
validation at $200{,}000$ shots with the same tables: $29$ of $33$ cells
fall inside the analytic bracket; the four outliers are low-count
statistics (e.g., the three correlated $60$-qubit cells at
$p{=}2\times10^{-3}$ observe $3$ failures against ${\sim}0.6$ expected; an
independent-seed $10^{6}$-shot recheck of evolved-H6 gives
$\pL=6.0\times10^{-6}$ with CI $[2.8\times10^{-6},1.3\times10^{-5}]$,
overlapping the bracket $[2.92,3.07]\times10^{-6}$). At $p{=}10^{-2}$,
where statistics are high, every cell agrees.

\begin{table*}[t]
\centering
\caption{Exact per-weight failure profile under the finite-weight
minimum-weight decoder tables (strict semantics). $f_w$ = failing fraction of
weight-$w$ Paulis; $F_w$ = absolute failing count; $W$ = table depth.
$f_1=f_2=0$ exactly for every row. For the surface block, additionally
$f_5=3.44\times10^{-1}$, $f_6=5.16\times10^{-1}$; its zero-syndrome bucket
contains $160$ weight-5 logicals and none at weight $\le4$, pinning
$\dist=5$ exactly. Surface routes are
$1-(1-p_{\rm block})^{n_{\rm modes}}$.}
\label{tab:fw}
\begin{ruledtabular}
\begin{tabular}{lccccccc}
code & $n$ & $k$ & $W$ & $f_3$ & $f_4$ & $F_3$ & $F_4$ \\
\colrule
evolved-H4   & 48  & 7  & 4 & $1.31\times10^{-2}$ & $5.73\times10^{-2}$ & 6{,}122 & 903{,}232 \\
evolved-H6   & 60  & 11 & 4 & $1.06\times10^{-2}$ & $4.58\times10^{-2}$ & 9{,}818 & 1{,}808{,}535 \\
evolved-LiH  & 60  & 11 & 4 & $9.41\times10^{-3}$ & $4.09\times10^{-2}$ & 8{,}695 & 1{,}614{,}926 \\
evolved-BeH2 & 84  & 13 & 4 & $1.90\times10^{-3}$ & $8.44\times10^{-3}$ & 4{,}893 & 1{,}319{,}242 \\
evolved-H2O  & 84  & 13 & 4 & $1.90\times10^{-3}$ & $8.44\times10^{-3}$ & 4{,}893 & 1{,}319{,}242 \\
floor-H6     & 60  & 11 & 4 & $1.06\times10^{-2}$ & $4.58\times10^{-2}$ & 9{,}818 & 1{,}808{,}535 \\
floor-BeH2   & 70  & 13 & 4 & $3.59\times10^{-3}$ & $1.65\times10^{-2}$ & 5{,}310 & 1{,}223{,}389 \\
floor-H2O    & 70  & 13 & 4 & $3.59\times10^{-3}$ & $1.65\times10^{-2}$ & 5{,}310 & 1{,}223{,}389 \\
floor-NH3    & 80  & 15 & 4 & $2.42\times10^{-3}$ & $1.10\times10^{-2}$ & 5{,}361 & 1{,}404{,}265 \\
floor-N2     & 100 & 19 & 4 & $1.99\times10^{-3}$ & $8.57\times10^{-3}$ & 8{,}669 & 2{,}720{,}913 \\
surface $[[25,1,5]]$ & 25 & 1 & 6 & $4.45\times10^{-2}$ & $1.62\times10^{-1}$ & 2{,}762 & 166{,}276 \\
\end{tabular}
\end{ruledtabular}
\end{table*}

\section{Domain of validity of the discovered family}
\label{app:family}

This appendix details the certification work behind
\S\ref{sec:results-family}: the labeling-class census and the resulting
linear qubit floor. Every certificate is computed by the same exact
strict-coset verifier as the rest of the paper (App.~\ref{app:eval}).

\paragraph{The bare family and its labeling dependence.}
Strip the discovered rule to its skeleton: the simulation graph is the
complete graph $K_N$ minus a 2-factor, with plain distance-3 operators, the
identity mode assignment, no ancilla padding, and no interaction-aware
choice of which edges to delete. Whether this bare family certifies strict
$\dist\ge5$ turns out to depend not on the pair $(N,\text{cycle type})$ but
on the \emph{vertex labeling} of the deleted 2-factor---specifically,
whether successive cycle vertices alternate between the two spin halves
(\emph{spin-alternating}) or run through consecutive indices
(\emph{block-canonical}). A coarse early sweep that fixed block-canonical
labeling had suggested a parity law in the per-vertex qubit count
$q_v=\lceil(N{-}3)/2\rceil$; the exhaustive census below shows that pattern
is an artifact of that single labeling class, not a property of the cycle
type.

\paragraph{The labeling-class census.}
The census evaluates $281$ labeled family points across $N=10$--$22$,
exhausting every cycle type at each $N$ under block-canonical labeling and
probing the spin-\emph{alternating} representative (deleted-cycle vertices
alternating between the two spin halves, defined for all-even types) of
every all-even type, plus random permutations and semi-alternating probes
at $N{=}14$, and the three constructor-diagnosed $2$-factors (BeH$_2$ and
H$_2$O at $N{=}14$, NH$_3$ at $N{=}16$) rebuilt as bare-family members
(\S\ref{sec:results-family}); Table~\ref{tab:census} tabulates the $278$
systematic-sweep points. The findings are as follows. Certified distance is \emph{not} a function of $(N,$ cycle
type$)$: at $N{=}14$, type $6{+}4{+}4$, the block-canonical labeling and
$20$ random relabelings all fail while the alternating representative
certifies. At $N\equiv2\ (\mathrm{mod}\ 4)$, no block-canonical 2-factor of
\emph{any} cycle type certifies (exhaustive: $0$ of the $124$ types at
$N\in\{10,14,18,22\}$), while the alternating representative certifies for
\emph{every} all-even type at $N=14$--$22$, including the single
Hamiltonian cycle; a single same-half adjacency (semi-alternating) already
fails ($0/4$). $N{=}10$ fails in every class tested. At odd $q_v$
($N=12,16,20$), alternating representatives certify all-even types
$23/23$; block labelings certify $46/79$ type-dependently (every all-even
type certifies; every failure contains an odd cycle).

\begin{table}[h]
\centering
\caption{The labeling-class census: certifying fraction for strict
$\dist\ge5$ over the bare family $\mathrm{GSE}(K_N-F)$, by vertex-labeling
class of the deleted 2-factor $F$. ``Block'' = block-canonical labeling,
exhaustive over all cycle types at each $N$; ``alternating'' = the
spin-interleaved representative of each all-even type; ``random'' = random
vertex permutations of type $6{+}4{+}4$; ``semi-alt.'' = one same-half
adjacency per odd cycle. The four sampled classes shown total $278$ points;
with the three constructor-diagnosed $2$-factors (BeH$_2$/H$_2$O at $N{=}14$,
NH$_3$ at $N{=}16$; see \S\ref{sec:results-family}) the full census is $281$ points.}
\label{tab:census}
\begin{ruledtabular}
\begin{tabular}{cccccc}
$N$ & $q_v$ & block & alternating & random & semi-alt. \\
\colrule
10 & 4  & $0/5$   & $0/2$   & ---    & --- \\
12 & 5  & $6/9$   & $4/4$   & ---    & --- \\
14 & 6  & $0/13$  & $4/4$   & $0/20$ & $0/4$ \\
16 & 7  & $12/21$ & $7/7$   & ---    & --- \\
18 & 8  & $0/33$  & $8/8$   & ---    & --- \\
20 & 9  & $28/49$ & $12/12$ & ---    & --- \\
22 & 10 & $0/73$  & $14/14$ & ---    & --- \\
\end{tabular}
\end{ruledtabular}
\end{table}

\paragraph{Minimum-weight logical and the linear floor.}
In a GSE encoding the mode-parity (vertex) operators are logical operators
of weight $\lceil v/2\rceil$ on a valence-$v$ active vertex, so the bare
family carries an a priori distance cap $\dist\le\lceil(N{-}3)/2\rceil$.
The exact-distance enumerations (App.~\ref{app:semantics}) show the
minimum-weight logicals of the $48$--$60$-qubit codes are exactly such
weight-5 vertex operators; for the larger codes lighter structural
logicals (still weight 5, or 6 at the $180$-qubit point) take over the
minimum. The necessity direction holds at all sizes: since vertex
operators are logicals, holding $\dist{=}5$ requires active valence
$\ge9$, i.e.\ at least ${\sim}5$ qubits per mode---a linear floor for the
whole family, and the objective of \S\ref{sec:results-floor}. The dense
rule pays $N\lceil(N{-}3)/2\rceil\approx N^{2}/2$, so its overhead over
the floor grows with size: $1.0\times$ at $12$ modes, $1.4\times$ at $16$,
$1.8\times$ at $20$.

\section{Reproducibility and data provenance}
\label{app:provenance}

Every number in this paper is recomputed offline by an exact strict-coset
verifier (App.~\ref{app:eval}) that we treat as the system of record, with
qBraid's GSE implementation as the only nonstandard dependency; the
evolutionary search services are treated as \emph{sources}, not
authorities, and are never queried for a reported quantity. The apparatuses
are disclosed in \S\ref{sec:methods}. The discovery experiments---the
distance climb and the fixed-distance compression runs, including the
$d{=}5$ run ($1411$ candidates, $272\to168$ panel qubits)---ran on Google's
AlphaEvolve Cloud API under an early-access program~\cite{novikov2025},
with mutations proposed by a $50/50$ mixture of Gemini 3.1 Pro and Gemini 3
Flash. The cross-apparatus replication, the floor-family run, the two
stabilizer-basis follow-ups, and the pre-submission controls all ran on the
open-source ShinkaEvolve~\cite{lange2025}, driven by GPT-5.5 or the cheaper
GPT-5.4 (Azure OpenAI) under hard per-run budget caps. Every certification
and validation campaign---the labeling-class census and constructor
diagnosis, the exact-distance enumerations, the exact-decoder
logical-failure sweep, the matched-compute random-search baseline, and the
exact stabilizer-basis-optimality certificates---involves no LLM and no
search: each is a deterministic, resumable verifier script. Total paid LLM
usage was under \$$250$ across all searches reported here (the AlphaEvolve
runs drew on early-access program quota); all verification and validation
ran on local Nvidia GH200 nodes.

All search configurations, seeds, evaluators, scoring files, program dumps,
and raw outputs are retained alongside the experiment records. This is not a
public reproduction package: the current GSE implementation and evaluator
are closed source, and the source snippets in this manuscript document the
discovered rules and the problem framing but do not run stand-alone outside
qBraid's GSE environment. All distance and logical-failure quantities use
\emph{strict} stabilizer-coset semantics throughout
(App.~\ref{app:semantics}).

\begin{table}[h]
\centering
\caption{Provenance map: each headline quantity and the experiment or
deterministic campaign whose logged records it derives from. Campaigns
marked ``no search'' involve no LLM; every quantity is recomputed by the
strict verifier in qBraid's internal GSE environment, GPU-bound sweeps on
GH200 nodes.}
\label{tab:provenance}
\scriptsize
\begin{ruledtabular}
\begin{tabular}{ll}
quantity & source \\
\colrule
distance-7 climb (\S\ref{sec:results-climb})        & climb run \\
distance-3 control (\S\ref{sec:results-efficiency}) & d-3 compression run \\
distance-5 winner / qubits (Table~\ref{tab:headline}) & d-5 compression run $+$ rediscovery \\
exact distances $+$ witnesses (Table~\ref{tab:constructionspecs}) & enumeration campaign (no search) \\
exact $\pL$ profile, ratios, MC (Table~\ref{tab:fw}, Fig.~\ref{fig:pl}) & exact-decoder campaign (no search) \\
held-out transfer (Table~\ref{tab:headline})        & d-5 head-to-head $+$ exact-decoder \\
conventional baselines (qubits)                     & baseline run \\
CGX non-isomorphism (\S\ref{sec:related})           & d-5 head-to-head \\
labeling-class census (Table~\ref{tab:census})      & census campaign (no search) \\
constructor diagnosis (\S\ref{sec:results-family}) & census $+$ post-hoc sweeps (no search) \\
large-molecule transfer (Table~\ref{tab:constructionspecs}) & post-hoc sweep (no search) \\
random-search baseline B2 (\S\ref{sec:results-climb}) & sampling campaign (no LLM) \\
stabilizer-basis reductions (\S\ref{sec:disc-limits}) & basis-search runs \\
joint graph$+$basis search (\S\ref{sec:disc-limits}) & joint-search run \\
floor family (Table~\ref{tab:floorfamily})          & floor-family run $+$ validation \\
\end{tabular}
\end{ruledtabular}
\end{table}

The exact logical-failure numbers are produced by a standalone NumPy
reimplementation of the App.~\ref{app:eval} methodology---BFS-built
minimum-weight tables, exhaustive per-weight classification, analytic
expansion with explicit truncation brackets, and $200{,}000$-shot
Monte-Carlo validation through the same tables on both routes---kept
independent of the search evaluator as a cross-check. The CGX
non-isomorphism certificate recomputes the evolved winner's qubit count and
stabilizer-group weight enumerator on each molecule and compares them to the
published Chen--Gorshkov--Xu distance-5 invariants; both are
Clifford-plus-permutation invariants, so any mismatch certifies
non-isomorphism without a Clifford search (\S\ref{sec:related}). The
complete program sources, logged metrics, and JSON outputs of every campaign
are retained alongside each experiment's records.

\clearpage

\bibliographystyle{apsrev4-2}
\bibliography{refs}

\end{document}